\documentclass[11pt,times]{emulateapj}

\usepackage{natbib}

 \newcommand{\be}{\begin{equation}}
 \newcommand{\ee}{\end{equation}}

 \newcommand{\n}[1]{$n_{\rm H_2}=10^#1$~cm$^{-3}$}

 \newcommand{\colds}[3]{$N({\rm #1})=#2\times10^{#3}$ cm$^{-2}$}

\begin{document}

\title{Gas-grain models for interstellar anion chemistry}

\author{M. A. Cordiner\altaffilmark{1} and S. B. Charnley}
\affil{Astrochemistry Laboratory and the Goddard Center for Astrobiology, Mailstop 691, NASA Goddard Space Flight Center, 8800 Greenbelt Road, Greenbelt, MD 20770, USA}
\email{martin.cordiner@nasa.gov}
\altaffiltext{1}{Institute for Astrophysics and Computational Sciences, The Catholic University of America, Washington, DC 20064, USA}

\begin{abstract}

Long-chain hydrocarbon anions C$_n$H$^-$ ($n=4, 6, 8$) have recently been found to be abundant in a variety of interstellar clouds. In order to explain their large abundances in the denser (prestellar/protostellar) environments, new chemical models are constructed that include gas-grain interactions. Models including accretion of gas-phase species onto dust grains and cosmic-ray-induced desorption of atoms are able to reproduce the observed anion-to-neutral ratios, as well as the absolute abundances of anionic and neutral carbon chains, with a reasonable degree of accuracy. Due to their destructive effects, the depletion of oxygen atoms onto dust results in substantially greater polyyne and anion abundances in high-density gas (with $n_{\rm H_2}\gtrsim10^5$~cm$^{-3}$). The large abundances of carbon-chain-bearing species observed in the envelopes of protostars such as L1527 can thus be explained without the need for warm carbon-chain chemistry. The C$_6$H$^-$ anion-to-neutral ratio is found to be most sensitive to the atomic O and H abundances and the electron density. Therefore, as a core evolves, falling atomic abundances and rising electron densities are found to result in increasing anion-to-neutral ratios. Inclusion of cosmic-ray desorption of atoms in high-density models delays freeze-out, which results in a more temporally-stable anion-to-neutral ratio, in better agreement with observations. Our models include reactions between oxygen atoms and carbon-chain anions to produce carbon-chain-oxide species C$_6$O, C$_7$O, HC$_6$O and HC$_7$O, the abundances of which depend on the assumed branching ratios for associative electron detachment.

\end{abstract}

\keywords{
astrochemistry -- ISM: abundances -- ISM: clouds -- ISM: molecules
}

\section{Introduction}

Hydrocarbon anions have been shown to be an important part of the molecular inventory of the Galaxy by their recent discoveries in the quiescent molecular clouds \object{TMC-1} \citep{mcc06} and \object{Lupus-1A} \citep{sak10}, the prestellar cores \object{L1544} and \object{L1512} and the envelopes of the low-mass protostars \object{L1527}, \object{L1251A} and \object{L1521F} \citep{sak07,gup09,cor11}. Previous chemical models have been able to reproduce with reasonable accuracy the observed abundances of C$_6$H$^-$ and C$_8$H$^-$ in TMC-1 and IRC+10216 \citep[\emph{e.g.}][]{mil07,rem07,cor08,wal09}, but not so well in the denser environment of L1527 \citep{har08}. The abundances of CN$^-$ and C$_3$N$^-$ observed in IRC+10216 \citep{tha08,agu10} closely match the predictions made by the chemical model of \citet{cor09}. These studies show that the basic gas-phase chemical processes involving molecular anions are fairly well understood. So far, however, interactions between negatively-charged gas-phase species and dust grain surfaces have been largely neglected in chemical models for anions. \citet{flo07} modelled the C$_6$H$^-$ to C$_6$H abundance ratio ([C$_6$H$^-$]/[C$_6$H])  in `typical' dark-cloud gas (with density $10^4$~cm$^{-3}$), using a chemical model accounting for the detailed calculation of charge balance, including gas, dust grain and PAH charge exchange, but did not consider the effects of depletion/freeze-out of molecules onto the grains. Gas-grain interactions are theorized to be of fundamental importance for the abundances of gas-phase molecules in dense interstellar clouds \citep[\emph{e.g.}][]{bro90}, and \citet{cor11} proposed that oxygen freeze-out could help to explain the large abundances of carbon-chain-bearing species and anions observed in low-mass protostellar envelopes and prestellar cores with densities $n_{\rm H_2}\sim10^5-10^6$~cm$^{-3}$. 

The total abundance of molecular anions is believed to be an important factor affecting the ionization balance of the interstellar medium (ISM) \citep[\emph{e.g.}][]{lep88,flo07,wak08}. Anion abundances are theorized to be sensitive to electron attachment and photodetachment rates, and may therefore provide a useful tool for the determination of electron densities and cosmic ray/photo- ionization rates in astrophysical environments \citep{mil07,flo07}. As shall become apparent from the present study, the interstellar C$_6$H$^-$ anion-to-neutral ratio can also be a tracer of the abundances of gas-phase atomic hydrogen, oxygen and carbon, due to the rapid reactions between hydrocarbon anions and these species.

This article presents new chemical models for interstellar carbon chains and their anions, with the aim of understanding the more recent C$_6$H$^-$ detections in relatively dense prestellar cores and cold protostellar envelopes. The inclusion of gas-grain interactions at varying levels of complexity permits analysis of the importance of these processes for anion chemistry, particularly at higher densities. We also discuss the possibility that reactions between hydrocarbon anions and oxygen atoms can produce detectable abundances of the carbon chain oxides C$_n$O and HC$_n$O ($n=6,7$) and the smaller hydrocarbon anion C$_2$H$^-$; these species are yet to be detected in space.

\section{Chemical modelling}
\label{sec:model}

The chemical models presented here utilize the (dipole-enhanced) RATE06 network \citep{woo07}, augmented with the anion chemistry of \citet{wal09}, and extended further to include C$_3$H$^-$ and C$_2$H$^-$.  Anions in the model are formed by radiative electron attachment, with rates taken from the calculations by \citet{her08}. Anions are destroyed in reactions with H, C, N and O atoms and in mutual-neutralisation reactions with the most abundant cations (including C$^+$, H$^+$, H$_3^+$ and HCO$^+$). Product branching ratios for anion-neutral reactions are taken from \citet{eic07}, with additional data from V. Bierbaum (private communication, 2010).  For reactions involving C$_3$H$^-$, the average was taken of the published rate coefficients for C$_2$H$^-$ and C$_4$H$^-$. In addition to the C$_3$O molecule already present in the RATE06 network, we have added the longer carbon-chain oxides C$_n$O, HC$_n$O (for $n=6$ and 7), as well as their protonated forms ((H)C$_n$OH$^+$) and anions ((H)C$_n$O$^-$) (see Section \ref{sec:cno} for details). PAHs have not been included in our models due to a lack of evidence concerning their abundances in dark molecular clouds \citep{pee11}.

The abundances of 454 gas-phase species are calculated as a function of time, linked by 5132 binary reactions. Models begin at time $t=0$ with neutral atomic abundances given in Table \ref{tab:parents}. The abundances of heavier metals (Na, Mg, Fe, P etc.) are set to zero, consistent with their non-detection in the ISM \citep{tur91}. The cosmic ray ionisation rate is set at $1.3\times10^{-17}$~s$^{-1}$ and the visual extinction ($A_V$) is set to 10 mag. Hydrogen is predominantly molecular and models are run at densities $n_{\rm H_2}=10^4$, 10$^5$ and 10$^6$~cm$^{-3}$. These span the range of interstellar cloud densities where anions have so far been observed, from quiescent, dense molecular clouds (\emph{e.g.} TMC-1, with \n{4}), to prestellar cores (\emph{e.g.} L1512, with \n{5}) and protostellar envelopes (\emph{e.g.} L1527, with \n{6}). The kinetic temperature is fixed at 10~K. These simplified models are useful for analysing basic astrochemical processes, and we do not consider the impact on the chemistry of protostellar heating, nor dynamical effects such as infall or outflows that may be present in some of the sources. Throughout this article, fractional abundances are given with respect to the total H$_2$ number density ($n_{\rm H_2}$).

\subsection{Model {\sc i}: No accretion}

Our first model considers only gas-phase chemistry and neglects the gas-grain interaction apart from the formation of H$_2$ on dust grains, which is assumed to occur at a rate $4.7\times10^{-18}n_{\rm H}(n_{\rm H}+2n_{\rm H_2})$. This modelling approach is the same as that used by \citet{har08} for L1527 and \citet{wal09} for TMC-1 and is therefore useful as a benchmark for comparison with those models. For details of the modelling procedure see \citet{woo07} and \citet{wal09}.

\subsection{Model {\sc ii}: Accretion without desorption}

This model builds on Model {\sc i} with the inclusion of gas-grain interactions. These are treated using the approach of \citet{cha97}, whereby gas-phase species (except for the highly volatile H$^+$, H$_2$, He and H$_3^+$) that collide with dust grains have a certain probability of sticking to the surface. We model a dust grain population composed of three distinct charge states: G$^0$ (neutral), G$^+$ (single positive charge) and G$^-$ (single negative charge). Due to electrostatic repulsion, anions are assumed not to stick to negatively-charged grains. Anions undergo mutual neutralisation upon collision with positively-charged grains, resulting in transfer of an electron from the anion (X$^-$) to the grain, and returning the neutral species (X) to the gas. Positive ions (X$^+$) undergo mutual-neutralisation in collisions with negatively-charged grains, again returning the neutral species (X) to the gas. The dust grain charges are calculated as a function of time using the formulation of \citet{ume90} for a uniform-size grain population, with the addition of the anion-grain processes mentioned above.  The parameters used are: dust grain radius $=10^{-5}$~cm, dust grain abundance $=2.0\times10^{-12}$, electron sticking-coefficient $=0.6$, gas sticking-coefficient $=1$, temperature $T=T_{\rm dust}=T_{\rm gas}$. The charge balance for this chemical plasma is

\begin{equation}
\sum_{\rm X^-} n({\rm X^-}) + n({\rm G}^-) + n_e  = \sum_{\rm X^+} n({\rm X}^+) + n({\rm G}^+)
\end{equation}

where the sums are over the number densities of all anions (X$^-$) and cations (X$^+$) in the model.

Due to the rapidity of collisions of electrons with the grains (as compared to cations with grains), the majority of the dust attains a negative charge soon after the start of the model, resulting in an approximately steady fraction of $\sim90$\% of the total number of grains as G$^-$ during the chemically-relevant timescales.  At a density of $10^6$~cm$^{-3}$, gas-grain collisions result in near-complete depletion of the gas after $\sim10^5$ yr (see Figure \ref{fig:n6}, middle panel).

\subsection{Model {\sc iii}: Accretion with desorption of atoms}

Gas-phase species deposited on the dust grains build up over time to form an icy mantle \citep[see for example][]{obe11}. Energetic processes are theorized to result in the liberation of mantle species back into the gas phase \citep{leg85}. Thermal desorption is very slow at dust temperatures of 10~K, and we assume that photo-desorption may be neglected in the dark clouds we are considering. However, heating of the dust grains by cosmic ray impact results in significantly elevated temperatures for short time periods, sufficient for volatile species to be quite rapidly desorbed \citep{has93}. Earlier models have proposed that atoms may either stick inefficiently on dust grains, or are selectively desorbed with respect to molecules \citep{has93,cha01,flo05}. In our final model, we utilize the rate coefficients for cosmic-ray-induced desorption of C, N, O and S atoms calculated by \cite{has93}. For simplicity, and due to their expected higher binding energies, desorption of molecular species is not considered.

\begin{deluxetable}{lc}
\tablewidth{0pt}
\tablecaption{Initial fractional abundances \label{tab:parents}}
\tablehead{
\colhead{Species} & \colhead{Abundance\tablenotemark{a}}} 
\startdata
H&1.0$\times10^{-3}$\\
He&2.8$\times10^{-1}$\\
C&1.2$\times10^{-4}$\\
N&4.0$\times10^{-5}$\\
O&3.2$\times10^{-4}$\\
S&2.0$\times10^{-8}$
\enddata
\tablenotetext{a}{Relative to $n_{\rm H_2}$.} 
\end{deluxetable}

\section{Results}
\label{sec:results}

\begin{figure}
\centering
\includegraphics[width=0.7\columnwidth,angle=270]{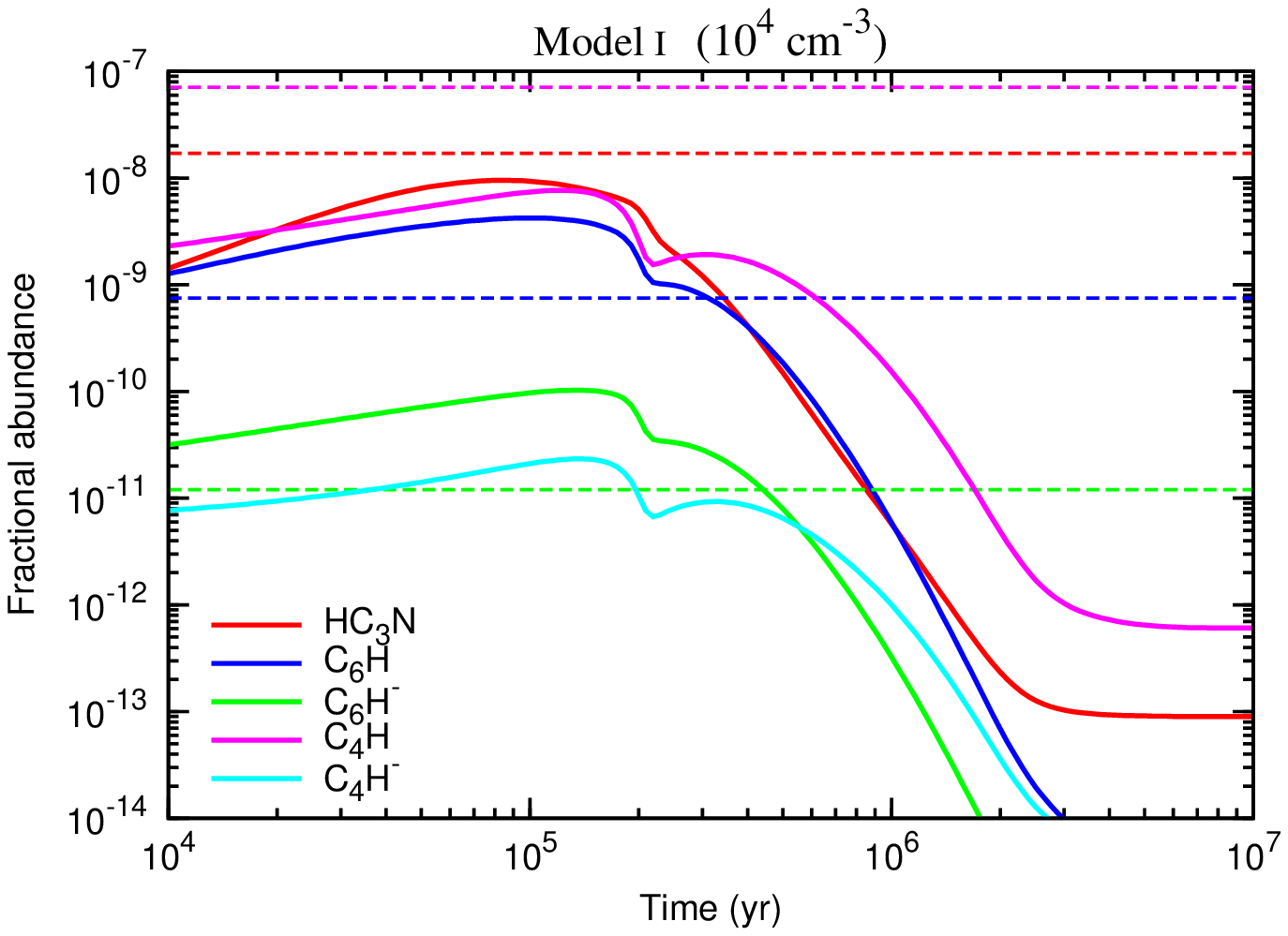}
\includegraphics[width=0.7\columnwidth,angle=270]{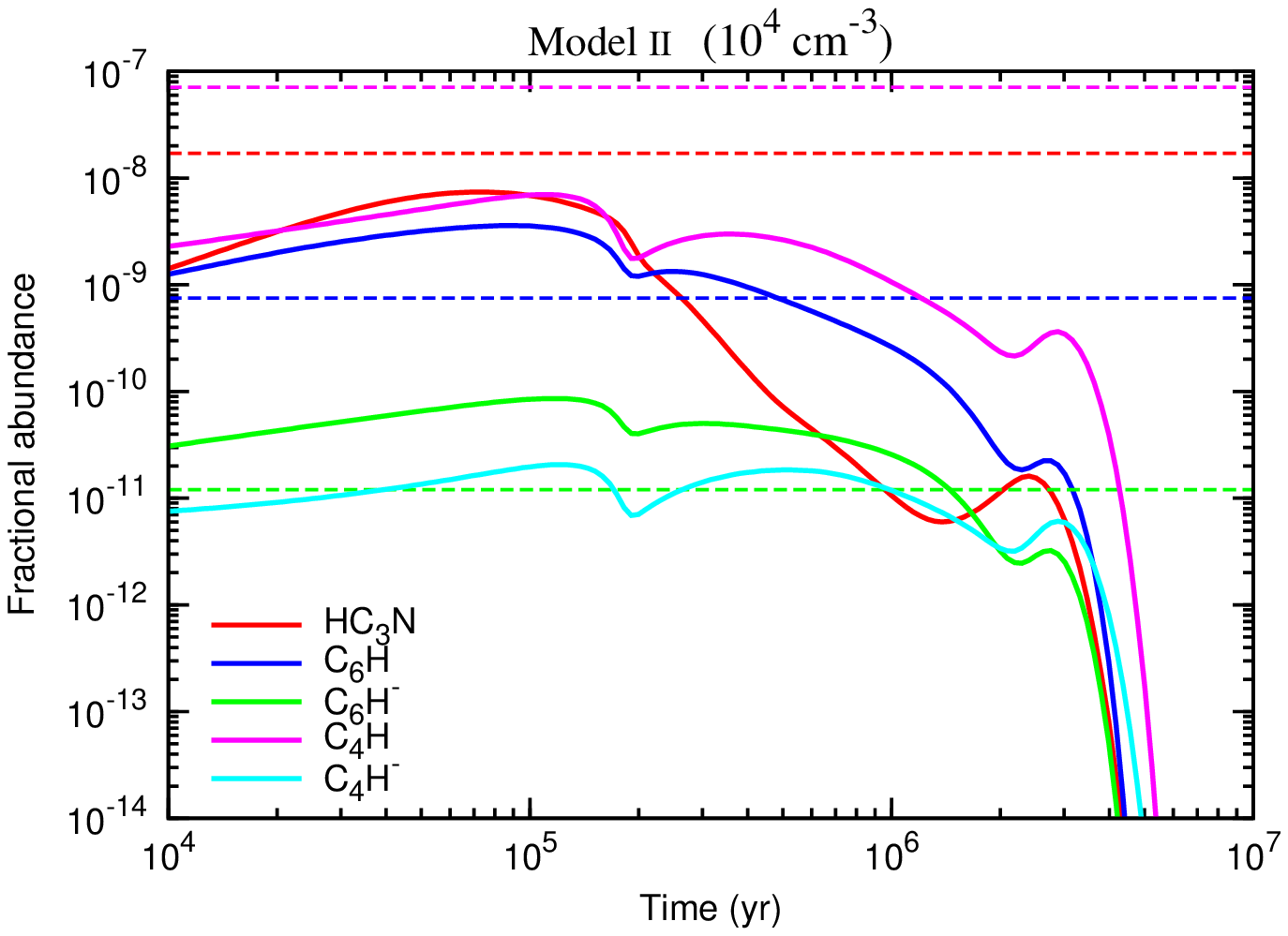}
\includegraphics[width=0.7\columnwidth,angle=270]{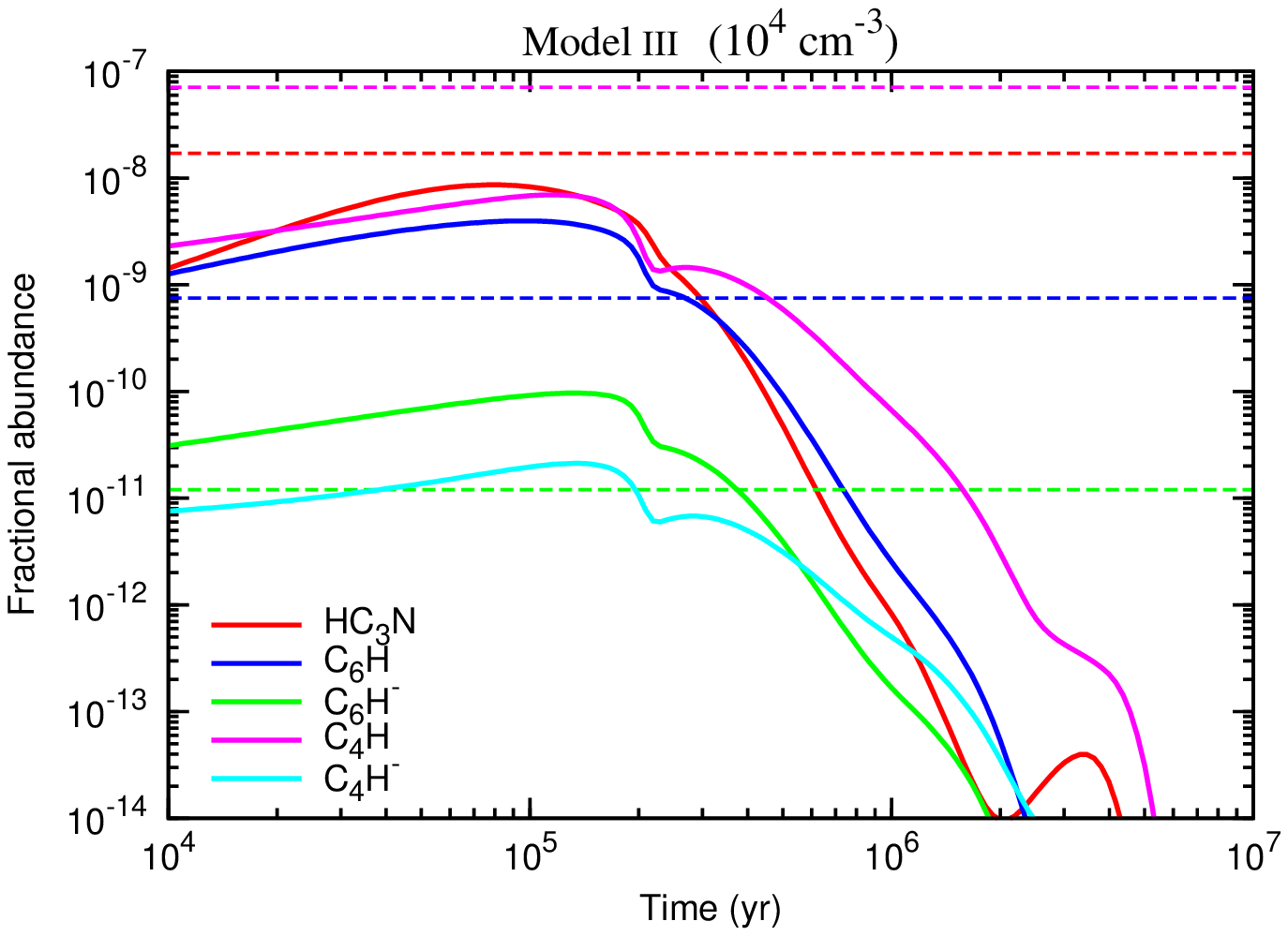}
\caption{Carbon chain and anion fractional abundances as a function of time for \n{4}. Results of Model {\sc i} (no accretion) are shown in the top panel,  Model {\sc ii} (accretion with no desorption) in the middle panel and Model {\sc iii} (accretion + CR-desorption) in the bottom panel. TMC-1 observed abundances are shown with dashed lines (data from \citealt{suz92,bru07,agu08}, with \colds{H_2}{1.0}{22} from \citealt{cer87}).}
\label{fig:n4}
\end{figure}

\begin{figure}
\centering
\includegraphics[width=0.7\columnwidth,angle=270]{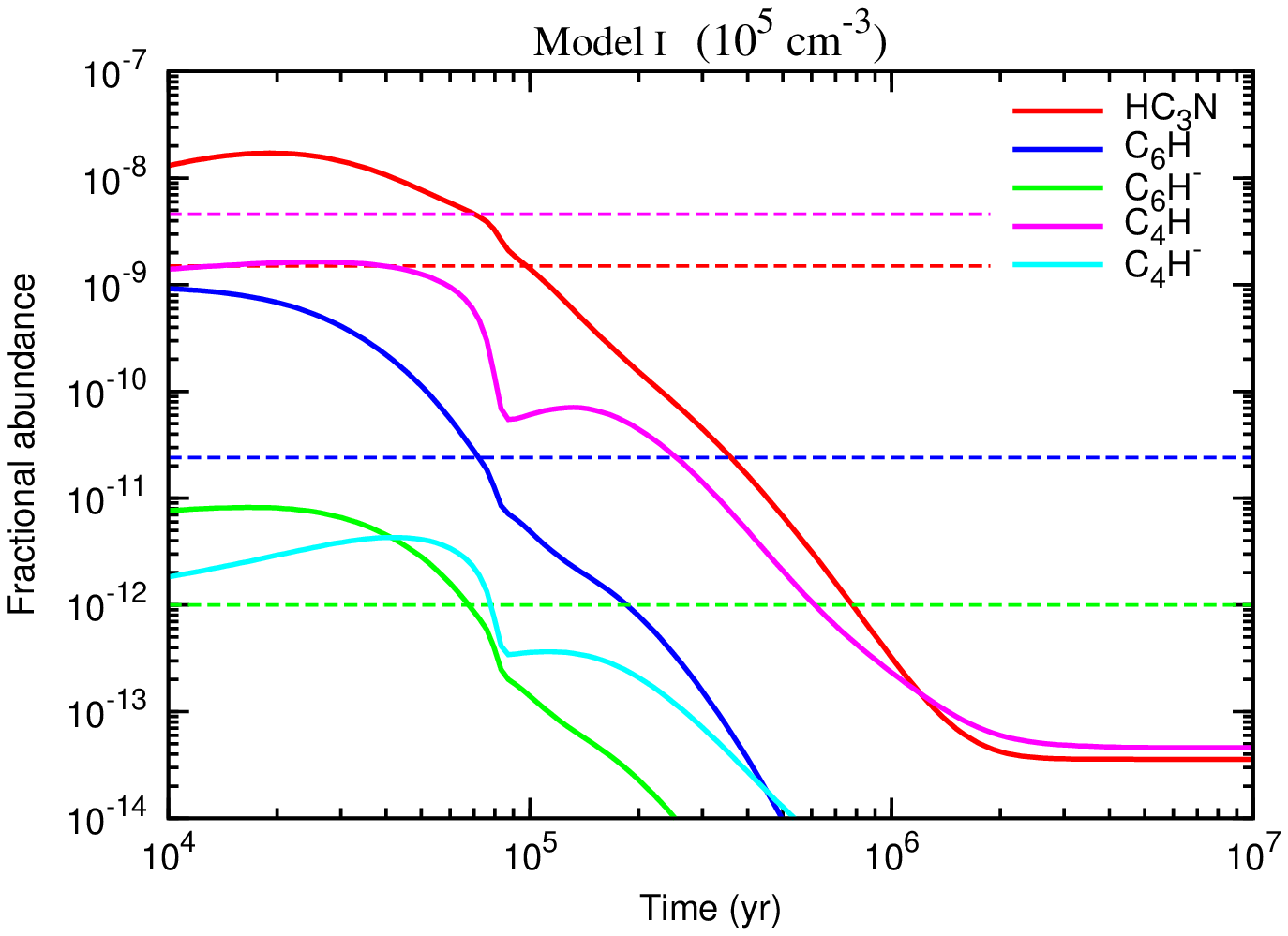}
\includegraphics[width=0.7\columnwidth,angle=270]{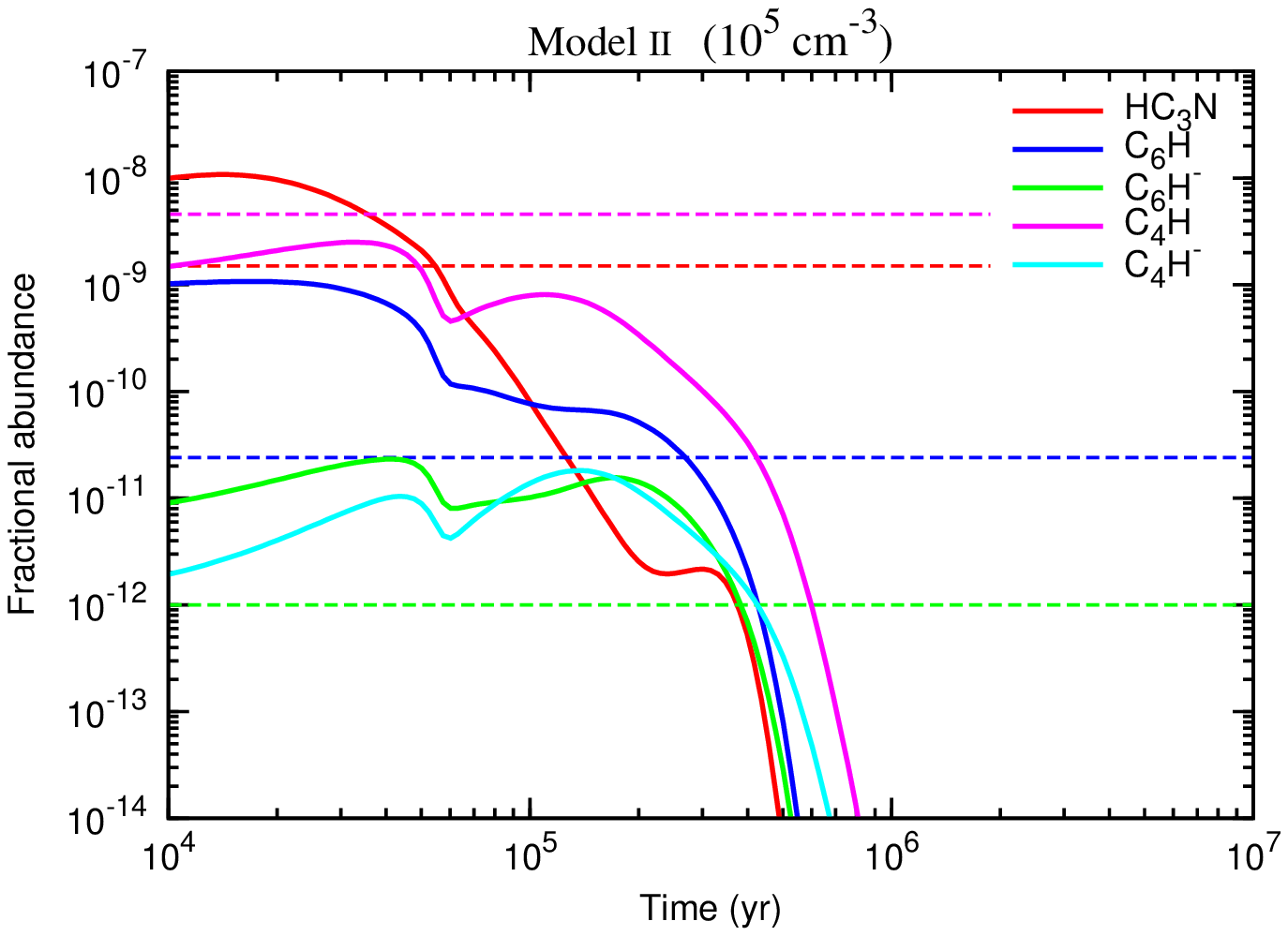}
\includegraphics[width=0.7\columnwidth,angle=270]{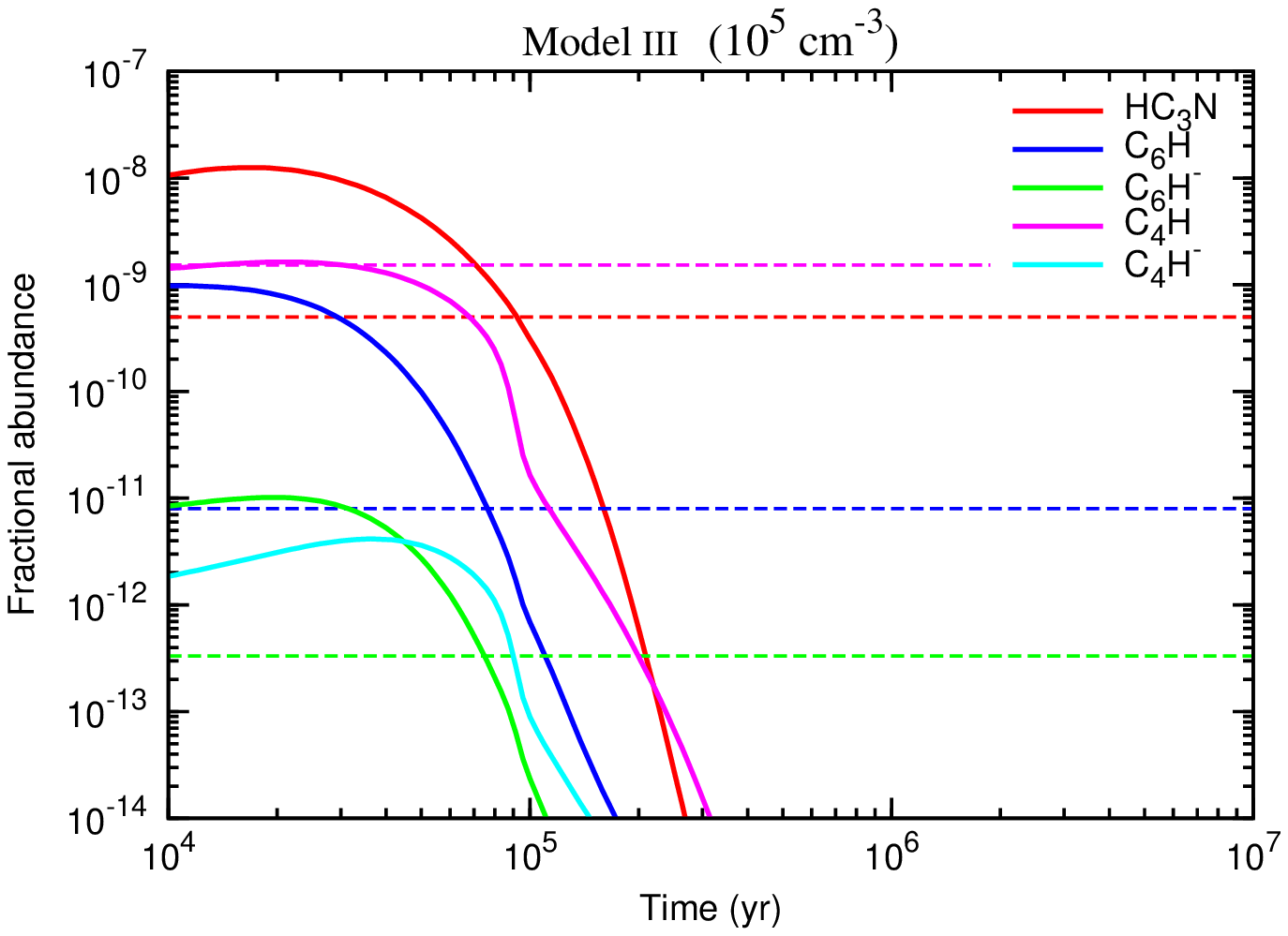}
\caption{As Figure \ref{fig:n4} but for \n{5}. L1512 observed abundances are shown with dashed lines (data from \citealt{cor11}, with \colds{H_2}{2.0}{22} from \citealt{kir05}).}
\label{fig:n5}
\end{figure}

\begin{figure}
\centering
\includegraphics[width=0.7\columnwidth,angle=270]{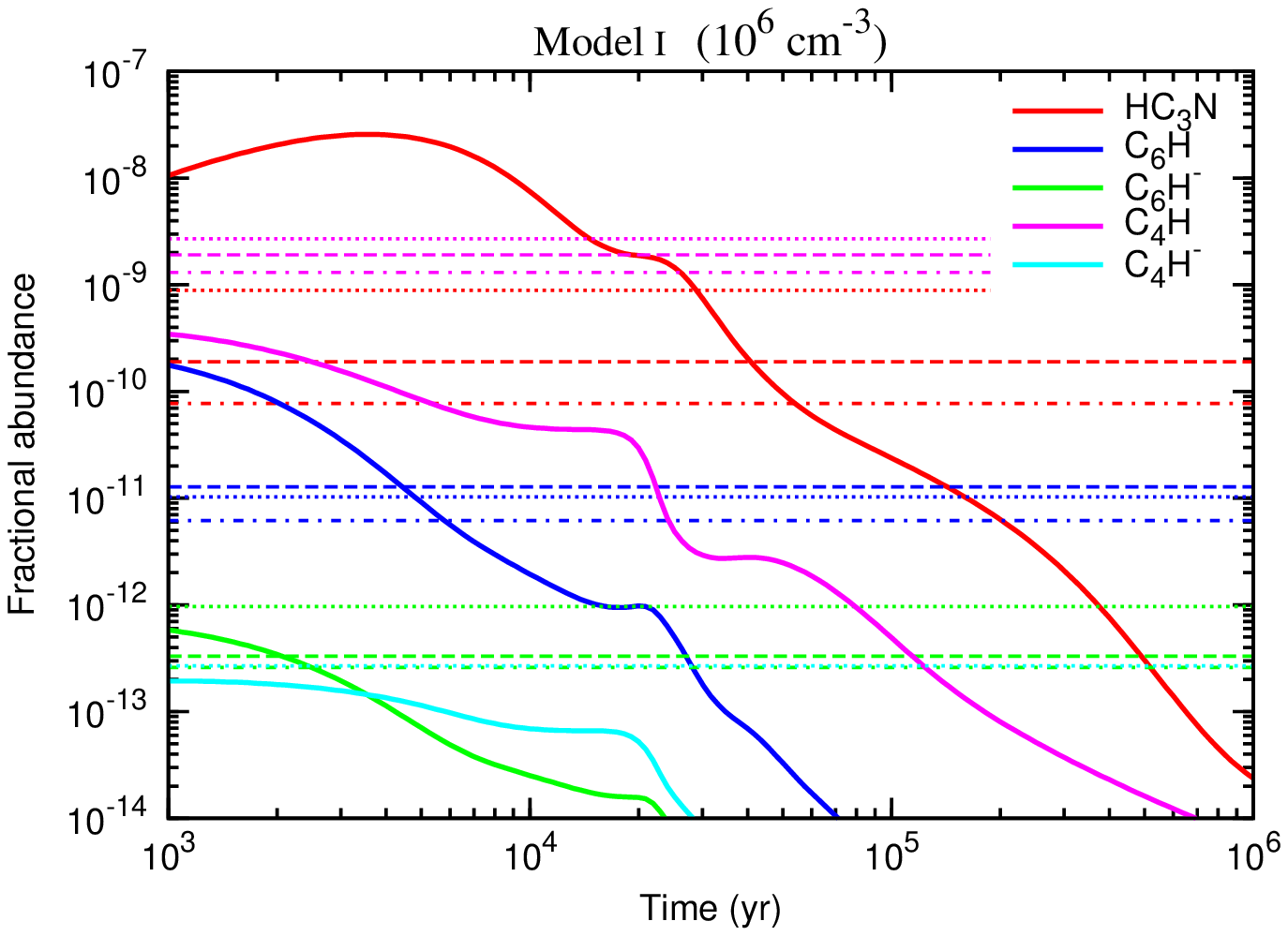}
\includegraphics[width=0.7\columnwidth,angle=270]{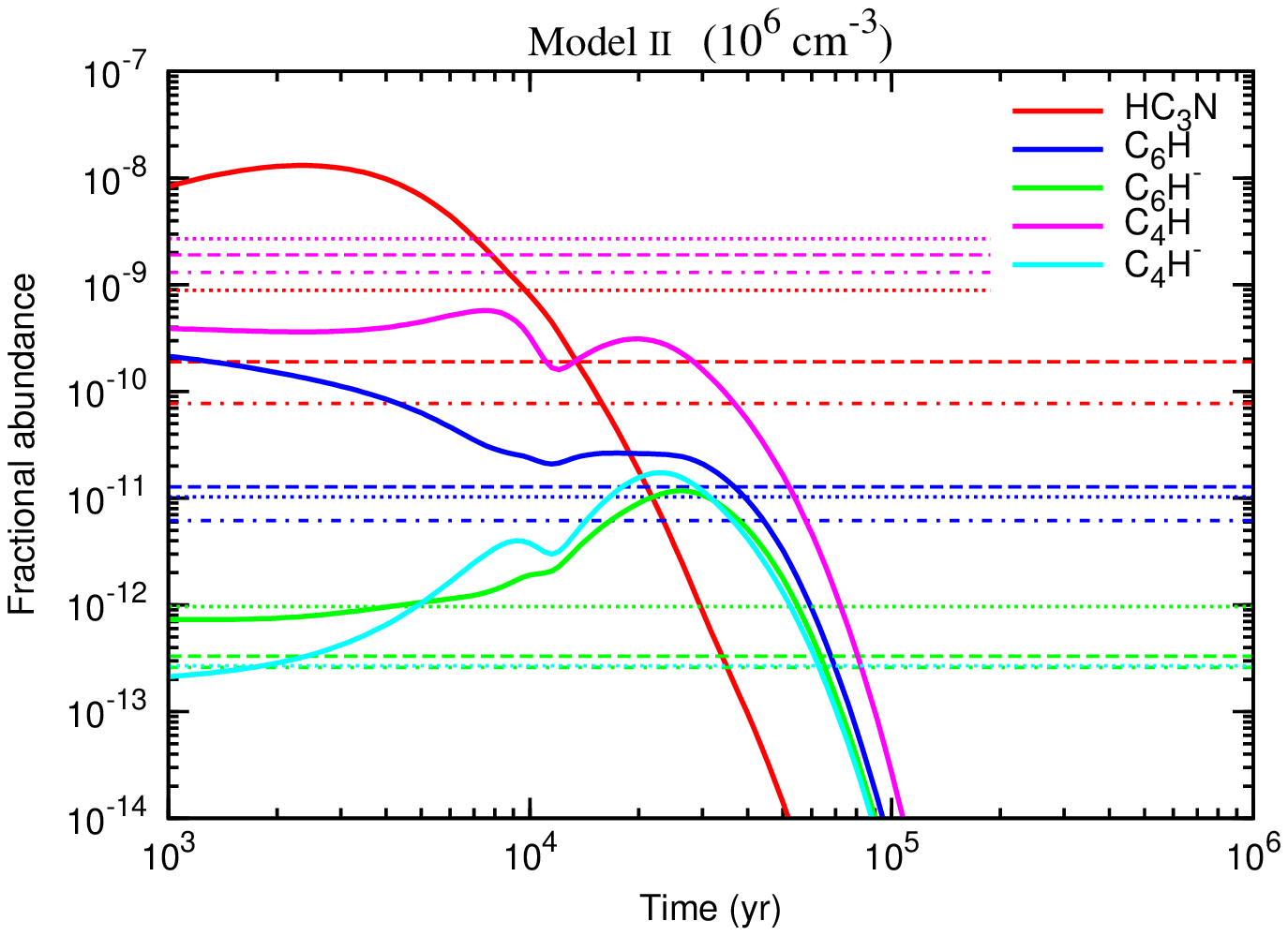}
\includegraphics[width=0.7\columnwidth,angle=270]{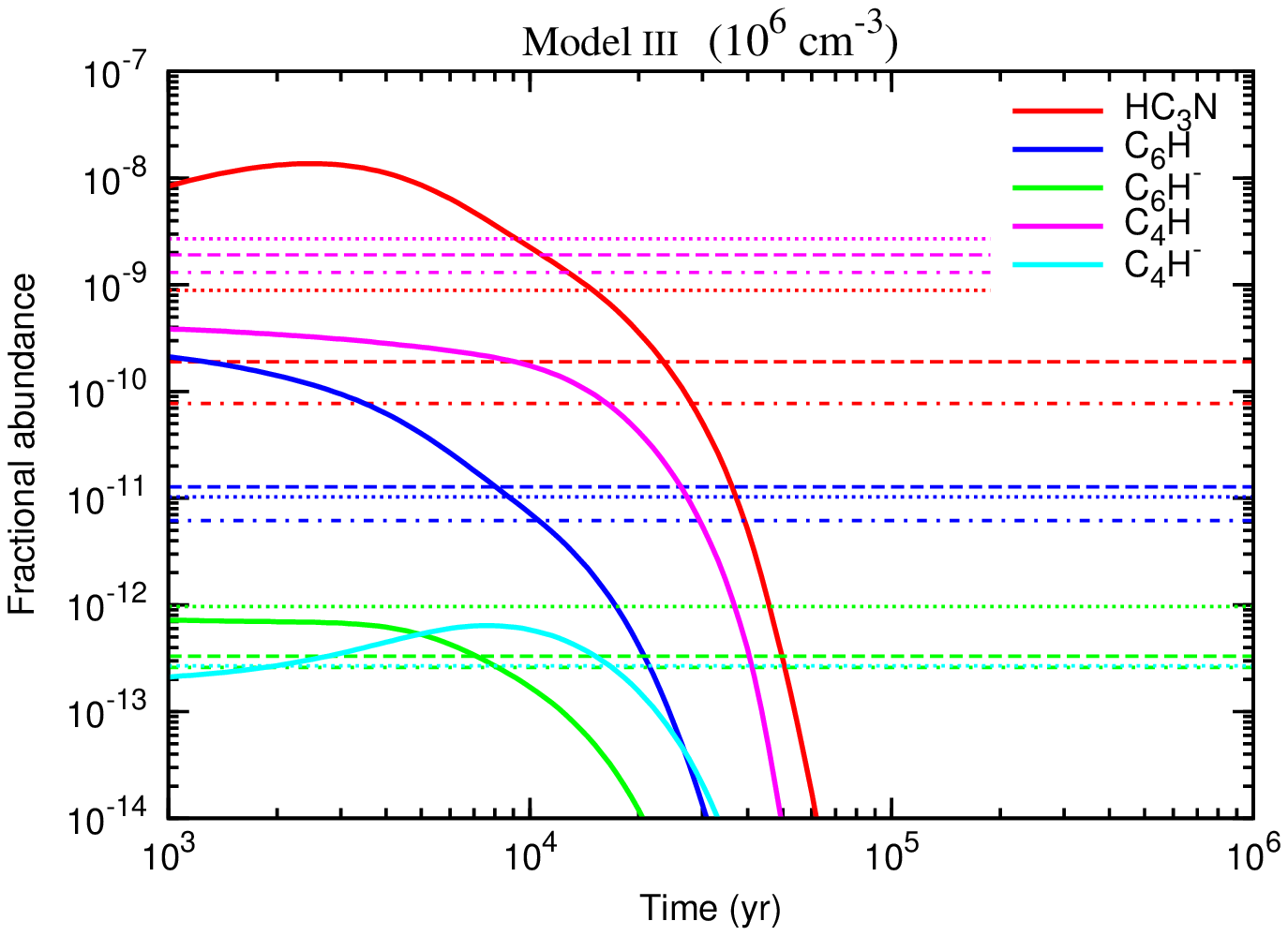}
\caption{As Figure \ref{fig:n4} but for \n{6}. L1544 abundances are shown with dashed lines (data from \citealt{suz92} and \citealt{gup09}, with \colds{H_2}{9.4}{22} from \citealt{cra05}). L1521F abundances are shown with dot-dashed lines (\citealt{gup09}, and our own unpublished Onsala 3~mm HC$_3$N data -- for details of the Onsala observations see \citealt{buc06}. We used \colds{H_2}{1.4}{23} from \citealt{cra05}). L1527 abundances plotted with dotted lines (\citealt{jor04,sak07,agu08}, with \colds{H_2}{9.4}{22} from \citealt{jor02}). The L1527 C$_4$H$^-$ abundance (cyan dotted line) is $2.7\times10^{-13}$ and lies just above the L1521F C$_6$H$^-$ abundance (green dot-dashed line).}
\label{fig:n6}
\end{figure}

For each of the three models, fractional abundances of the carbon-chain-bearing species HC$_3$N, C$_4$H, C$_6$H and anions C$_4$H$^-$ and C$_6$H$^-$ are plotted as a function of time in Figures \ref{fig:n4}-\ref{fig:n6} at densities $n_{\rm H_2}=10^4$, 10$^5$ and 10$^6$~cm$^{-3}$, respectively. Horizontal broken lines show the observed fractional abundances of each species: for \n{4}, TMC-1 data are shown and for \n{5}, L1512 data are shown. For \n{6}, observed abundances are plotted for three sources: the pre-stellar core L1544 (dashed lines), the very low-luminosity, young protostar L1521F (dotted lines), and the Class 0/I protostar L1527 (dot-dashed lines). We briefly examine the results obtained for each density regime.

\subsection{\n{4} -- Quiescent molecular cloud}

For \n{4} (Figure \ref{fig:n4}), Models {\sc i}-{\sc iii} achieve a reasonable match with observations of the quiescent molecular cloud TMC-1 (within about an order of magnitude) at times around 10$^5$~yr, consistent with previous models for this source \citep[\emph{e.g.}][]{her89,wal09}. At times later than this, the HC$_3$N and C$_4$H abundances fall well below their respective observed values. 

Interstellar polyyne (C$_n$H) and cyanopolyyne (HC$_n$N) chemistry was discussed by \citet{wal09}; polyynes are radical species and react rapidly with oxygen atoms whereas the closed-shell cyanopolyynes do not. Thus, in Model {\sc ii} the freezing-out of atomic oxygen onto dust grains eliminates one of the main polyyne destruction channels which significantly extends their lifetime in the gas. HC$_3$N, on the other hand, is primarily destroyed by reaction with C$^+$ and continues to fall in a similar manner between $10^5$ and $10^6$~yr independently of whether \mbox{O-atoms} are accreting out (see Figure \ref{fig:n4}, upper and middle panels). The cosmic-ray-desorption of \mbox{O-atoms} that occurs in Model {\sc iii} (Figure \ref{fig:n4}, lower panel) results in similar abundances to Model {\sc i} because at this relatively low density the gas accretion rate is sufficiently low that it is counteracted by cosmic-ray desorption, which keeps the majority of material off the dust and in the gas-phase.

At a time of $10^5$~yr, the modelled C$_6$H$^-$ anion-to-neutral ratio [C$_6$H$^-$]/[C$_6$H] is equal to 2.3\% in all three models with \n{4} and is thus apparently quite insensitive to gas-grain interactions at this time. The anion-to-neutral ratio rises at later times, predominantly as a result of rising electron densities and falling atomic H, C and O abundances (see Section \ref{sec:anr}). Comparison of the modelled anion-to-neutral ratio with the observed values given in Table \ref{tab:obs} shows good agreement with the quiescent clouds TMC-1 and Lupus~1A.

\begin{deluxetable}{lcccc}
\tablewidth{0pt}
\tablecaption{Observed anion-to-neutral ratios, densities and CO depletion factors in dense cloud cores \label{tab:obs}}
\tablehead{Source & [C$_6$H$^-$]/[C$_6$H]& $n_{\rm H_2}$ (cm$^{-3}$) & $\delta$(CO) & Refs.}
\startdata
TMC-1  & $1.6\pm0.3$\% & $\sim10^4$ & 1.2\tablenotemark{a} & 1, 2, 3\\ 
Lupus-1A&$2.1\pm0.6$\% & \nodata & \nodata & 4 \\
L1512  & $4.2\pm1.4$\% & $\sim10^5$ & 3.9\tablenotemark{b} & 5, 6, 7 \\
L1251A & $3.6\pm1.3$\% & $\sim10^5$ & \nodata & 5, 8\\
L1544  & $2.5\pm0.8$\% & $\sim10^6$ & 7-14 & 9, 10, 11\\
L1521F & $4.0\pm1.0$\% & $\sim10^6$ & 12-15 & 9, 10, 12\\
L1527  & $9.3\pm2.9$\% & $\sim10^6$ & 2.4\tablenotemark{a} & 13, 14
\enddata
\tablenotetext{a}{Calculated from CO observations assuming an undepleted abundance of $9.5\times10^{-5}n_{\rm H_2}$.} 
\tablenotetext{b}{The \citet{lee03} CO depletion factor in L1512 has been scaled by 0.35 to account for the greater undepleted CO  abundance of $2.7\times10^{-4}n_{\rm H_2}$ used in that study.} 
\tablerefs{(1) -- \citet{bru07}; (2) -- \citet{hir92}; (3) -- \citet{ohi92}; (4) -- \citet{sak10}; (5) -- \citet{cor11}; (6) -- \citet{kir05}; (7) -- \citet{lee03}; (8) -- \citet{lee10}; (9) -- \citet{gup09}; (10) -- \citet{cra05}; (11) -- \citet{cas02}; (12) -- \citet{cra04}; (13) -- \citet{sak07}; (14) -- \citet{jor04}. }
\end{deluxetable}

\subsection{\n{5} -- Prestellar core}

At \n{5} (Figure \ref{fig:n5}), a reasonable agreement is obtained with observations of the prestellar core L1512 at times around (6-8)$\times10^4$~yr. The polyynes and anions in Models {\sc i} and {\sc iii} (top and bottom panels) show a good fit to observations at this time (within a factor of a few), but HC$_3$N is significantly over-predicted. On the other hand, in Model {\sc ii} (with freeze-out and no desorption), HC$_3$N matches observations at about $6\times10^4$~yr whereas C$_6$H and C$_6$H$^-$ are over-predicted by about an order of magnitude; the abundances of the latter are greater as a result of \mbox{O-atom} freeze-out.

The best fit to the observed C$_6$H$^-$ anion-to-neutral ratios in the sources with \n{5} (L1512 and L1251A; see Table \ref{tab:obs}) is also obtained in Models {\sc i} and {\sc iii}, at around $7\times10^4$~yr, at which time [C$_6$H$^-$]/[C$_6$H]~=~3.1\% and 3.5\% respectively. The freeze-out model with no desorption (Model {\sc ii}), significantly over-predicts the anion-to-neutral ratio at this time ([C$_6$H$^-$]/[C$_6$H]~=~8\%) due to the larger ratiative electron attachment rate that results from the greater electron density that occurs during freeze-out (see Section \ref{sec:anr}).

\subsection{\n{6} -- Dense prestellar core / cold protostellar envelope}

Gas-grain interactions are more rapid at higher densities, so at \n{6} the effects of accretion on the modelled abundances are quite significant, as can be seen by comparison of the three panels in Figure \ref{fig:n6}. Gas-phase chemistry also proceeds more rapidly at high densities: without accretion (top panel), the conversion of C into CO (through the reaction of C and O$_2$), and the presence of abundant gas-phase \mbox{O-atoms} that react with and destroy the polyynes results in steadily diminishing carbon chain abundances over time. Accordingly, the Model {\sc i} abundances fail to match observations of L1544, L1521F or L1527 at any time.

The inclusion of gas-grain interactions results in substantially greater carbon chain abundances for $t\gtrsim1000$~yr in Model {\sc ii} compared with Model {\sc i}. Indeed, Model {\sc ii} shows a good fit (within an order of magnitude), to the observed abundances in L1544, L1521F and L1527 at $t\sim10^4$~yr. The only species that does not fit well with observations is C$_4$H$^-$, which has been over-predicted by at least an order of magnitude in all previous chemical models for this species \citep{mil07,flo07,har08,wal09,cor09}. The most likely explanation for this discrepancy is a problem with the theoretical radiative electron attachment rate for C$_4$H, which was calculated using a relatively simplistic phase-space approach \citep[for further discussion see][]{her08}.

The inclusion of cosmic-ray-induced desorption (Model {\sc iii}) diminishes the polyyne and anion abundances somewhat compared with Model {\sc ii}, primarily due to the increased gas-phase abundance of destructive \mbox{O-atoms.} The peak that occurs in the polyyne abundances in Model {\sc ii} towards the time of freeze-out is eliminated, but a good fit to observations is still achieved at $\sim10^4$~yr, although HC$_3$N is again somewhat over-abundant. The C$_6$H$^-$ anion-to-neutral ratio is found to be quite sensitive to the effects of gas-grain interactions at this density; in the absence of accretion (Model {\sc i}), [C$_6$H$^-$]/[C$_6$H]~=~1.3\% at $10^4$~yr (significantly less than observed values; see Table \ref{tab:obs}). With accretion and no desorption (Model {\sc ii}), [C$_6$H$^-$]/[C$_6$H]~=~8\% and rises sharply as time goes on (see Figure \ref{fig:andens}). When cosmic-ray-induced desorption of atoms is accounted for (Model {\sc iii}), [C$_6$H$^-$]/[C$_6$H] levels off after $\sim10^4$~yr at about 3\%, which matches well with the observations of L1544 and L1521F, but somewhat less well with the greater value of 9.3\% in L1527.

Dissociative electron attachment (DEA) to the linear carbene C$_6$H$_2$ has been suggested as a potentially important route to the formation of C$_6$H$^-$ in L1527 \citep{sak07}. This reaction was shown to be likely to proceed rapidly in the ISM by the theoretical calculations of \citet{her08}. In our models, C$_6$H$_2$ is included as the more stable poly-acetylene isomer, and reactions leading to the carbene(s) are neglected due to a lack of laboratory data. The observed (linear carbene) C$_6$H$_2$ abundances in TMC-1 and L1527 are reproduced if about 1-2\% of the total C$_6$H$_2$ in our models is assumed to be in the form of the carbene isomer. If the poly-acetylene to carbene ratio is fixed at 1\%, inclusion of the reaction 

\begin{equation}
{\rm C_6H_2} + e^- \longrightarrow {\rm C_6H^-} + {\rm H} 
\end{equation}

with a rate coefficient of 4.98$\times10^{-7}(T/300)^{-0.5}$ cm$^3$\,s$^{-1}$ \citep{her08}, generally raises the C$_6$H$^-$ abundances in our models by only a negligibly small amount. 

C$_6$H$_2$ is destroyed predominantly by reacting with atomic carbon to produce C$_7$H + H. Its abundance is therefore sensitive to the gas-phase \mbox{C-atom} abundance. At \n{5} in Model {\sc ii} (where C atoms are less abundant than in any of the other models), the C$_6$H$_2$ abundance can become sufficiently large ($1.6\times10^{-9}$ at $7\times10^4$~yr) that DEA becomes significant, contributing about 50\% to the total C$_6$H$^-$ formation rate. Observations of C$_6$H$_2$ in L1512 and other clouds of similar density will be required to further examine the possible significance of DEA for interstellar anion production. DEA is not expected to proceed for carbenes smaller than C$_6$H$_2$ because the reaction becomes endothermic, with an energy barrier exceeding the electron temperature in dense interstellar clouds \citep{her08}.  

\section{C$_2$H$^-$}
\label{sec:c2h}

The C$_2$H$^-$ radiative electron attachment rate is calculated to be very small ($2\times10^{-15}(T/300)^{-0.5}$ cm$^3$\,s$^{-1}$; \citealt{her08}). As such, little C$_2$H$^-$ has previously been expected in the ISM. However, the dominant production mechanism in our models is \emph{via} the rapid reaction 

\begin{equation}
{\rm C_3H^-} + {\rm O} \longrightarrow {\rm C_2H^-} + {\rm CO}
\end{equation} 

the rate of which is derived from interpolation of the C$_4$H$^-$ + O and C$_2$H$^-$ + O rates measured by \citet{eic07} (and divided by two to account for the 50\% branching ratio adopted for the associative electron detachment channel). 

In Model {\sc iii}, the calculated C$_2$H$^-$ fractional abundance is 6.6$\times10^{-14}$ at $t=10^5$~yr for \n{4} and 5.9$\times10^{-14}$ at $t=10^4$~yr for \n{6}. These abundances are low, but may be sufficient to permit detection with current or future radio telescopes. \citet{agu08} obtained upper limits on the C$_2$H$^-$ column densities in TMC-1 and L1527 of $2.2\times10^{11}$ and $1.8\times10^{10}$~cm$^{-2}$ respectively, which correspond to fractional abundance upper limits of $2.2\times10^{-11}$ and $3.0\times10^{-13}$. A deeper search for C$_2$H$^-$ in TMC-1 and L1527 would thus provide a useful test of our theory with regard to anion reactions with \mbox{O-atoms.}

\section{Carbon chain oxides}
\label{sec:cno}

In the laboratory experiments of \citet{eic07} on the rate of reaction of atomic oxygen with C$_n$(H)$^-$, a loss of ion signal was observed over time (V. Bierbaum and Y. Zhibo, private communication, 2010). This effect is attributable to associative electron detachment (AED) reactions 

\begin{equation}
{\rm O} + {\rm C}_n{\rm (H)}^- \longrightarrow {\rm (H)C}_n{\rm O} + e^- 
\end{equation}

The raw, unpublished laboratory data are consistent with a branching ratio for AED of approximately 0.5. It must be emphasised, however, that AED products were not directly observed, and the existence of the implied carbon-chain-oxide product channels can only be definitively confirmed by additional dedicated laboratory studies. We assume the alternative product channel (with braching ratio 0.5) to result in formation of a CO molecule and a smaller anion (as described in the previous section on C$_2$H$^-$):

\begin{equation}
{\rm O} + {\rm C}_n{\rm (H)}^- \longrightarrow {\rm (H)C}_{n-1}^- + {\rm CO} 
\end{equation}

Model calculations were performed using an AED branching ratio of 0.5 in the reaction of oxygen with the most abundant anions in the model: C$_6^-$, C$_7^-$, C$_6$H$^-$ and C$_7$H$^-$, to produce C$_6$O, C$_7$O, HC$_6$O and HC$_7$O, respectively. In addition to the new carbon-chain-oxide neutrals, their anions, cations and protonated forms were also added to the model. Radiative electron attachment rates for C$_6$O and C$_7$O were calculated assuming immediate formation of the stable anion (through vibronic transitions), following s-wave electron capture based on the method of \citet{her08}. The degeneracies of the ground states of C$_n$O for $n$ odd and even are 1 and 3 respectively, and 4 for the C$_n$O$^-$ anions \citep[based on the electronic structure calculations of][]{rie00}, which leads to radiative electron attachment rates of $3.26\times10^{-7}(T/300)^{-0.5}$~cm$^{3}$\,s$^{-1}$ for C$_6$O and $9.78\times10^{-7}(T/300)^{-0.5}$~cm$^{3}$\,s$^{-1}$ for C$_7$O. In the absence of detailed electronic structure information for the HC$_6$O$^-$ and HC$_7$O$^-$ anions, an approximate radiative attachment rate coefficient of $10^{-7}(T/300)^{-0.5}$~cm$^{3}$\,s$^{-1}$ has been used. 

Based on the reactions and rate coefficients of the structurally similar cumulene oxide C$_3$O in the RATE06 database, the new carbon-chain-oxides have been assumed to to undergo charge transfer reactions with H$^+$ and C$^+$, dissociative charge-transfer with He$^+$ and C$^+$ and protonation by H$_3$O$^+$ and HCO$^+$. Their cations have been assumed to react with H$_2$ and to undergo dissociative electron recombination. For the recombination of HC$_n$O$^+$ and H$_2$C$_n$O$^+$, an H atom is assumed to be ejected, whereas for C$_n$O$^+$, a CO molecule is assumed to be ejected. Due to the low fractional abundances of these ions ($\sim10^{-12}$), assuming alternative recombination products (\emph{e.g.} HC$_n$O$^+$ + $e^-~\longrightarrow~$C$_{(n-1)}$H + CO), has a negligible effect on the abundances of the species of interest in this study. Reactions of (H)C$_n$O$^-$ with H and C atoms (modelled using generic rate coefficients of $10^{-9}$~cm$^{3}$\,s$^{-1}$), are assumed to result in associative detachment \citep[as observed in the laboratory for hydrocarbon chain anions by][]{eic07}; their reaction with \mbox{O-atoms} is assumed to result in the production of a CO$_2$ molecule and an anion. The (H)C$_n$O$^-$ anion destruction mechanisms are through photodissociation and mutual neutralisation with the most abundant cations at a rate $7.51\times10^{-7}(T/300)^{-0.5}$~cm$^{3}$\,s$^{-1}$ \citep{wal09}. Thus, although additional reactions are likely, we anticipate that the most rapid processes involving these new carbon-chain-oxide species have been -- at least approximately -- accounted for. 

\begin{figure}
\begin{center}
\includegraphics[width=0.7\columnwidth,angle=270]{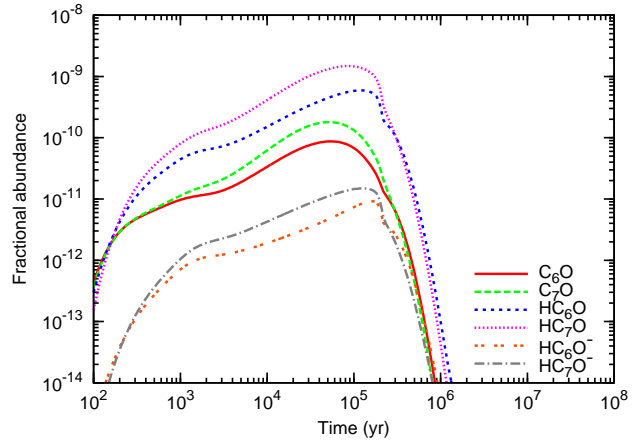}
\end{center}
\caption{Modelled carbon-chain-oxide fractional abundances as a function of time from Model {\sc iii} at \n{4}.}
\label{fig:TMC1CnO}
\end{figure}

The abundances of the carbon chain oxides C$_7$O, HC$_7$O and their anions (calculated at \n{4} using Model {\sc iii}), are plotted as a function of time in Figure \ref{fig:TMC1CnO}. The calculated abundances of these modelled oxides and their anions are given for densities applicable to TMC-1 and L1544 in Table \ref{tab:cno}. Around the time of best agreement of the \n{4} results with observations of TMC-1 ($\sim10^5$~yr), the C$_6$O and C$_7$O abundances are $6.8\times10^{-11}$ and $1.3\times10^{-10}$ respectively. The HC$_6$O and HC$_7$O abundances are $5.8\times10^{-10}$ and $1.5\times10^{-9}$ respectively. These are comparable to the C$_3$O abundance observed in TMC-1 by \citet{bro85} of $1.4\times10^{-10}$, which suggests that large carbon chain oxides may be detectable in TMC-1 and similar carbon-chain-rich clouds. The predicted carbon-chain-oxide abundances depend on the availability of gas-phase \mbox{O-atoms} and are therefore significantly less in the higher-density environment of L1544 as a result of depletion of \mbox{O-atoms} onto dust.

The most important route to the smaller cumulene oxide C$_3$O in our model is through the dissociative recombination of H$_3$C$_3$O$^+$, which is formed from the reaction of C$_2$H$_3^+$ with CO. We do not consider the equivalent mechanisms for C$_6$O or C$_7$O synthesis (\emph{via} C$_5$H$_3^+$ and C$_6$H$_3^+$, respectively), because these ions are much less abundant in the model than C$_2$H$_3^+$, especially at early times. 

The addition of carbon-chain oxides to our models has a relatively minor impact on the abundances of other species. They act to boost the abundances of long carbon chains and anions by a few percent due to a recycling effect, for example:

\begin{equation}
{\rm C_6H^- + O} \longrightarrow {\rm HC_6O} + e^- 
\end{equation}

followed by 

\begin{equation}
{\rm HC_6O + C} \longrightarrow {\rm C_6H + CO}
\end{equation}

whereas in previous models \citep[\emph{e.g.}][]{wal09}, \mbox{O-atom} reactions resulted only in the shortening of carbon chains through CO formation:

\begin{equation}
{\rm C_6H^- + O} \longrightarrow {\rm C_5H^- + CO}
\end{equation}

Further dedicated laboratory studies are required in order to firmly establish the products of reactions between anions and oxygen atoms. In addition, observational searches for carbon-chain oxides and their anions will help to constrain the relevant branching ratios. The neutral (H)C$_n$O species (for $n=6,7$) have known rotational spectra and possess large dipole moments, making them suitable candidates for radio-astronomical searches.

\begin{deluxetable}{lcc}
\tablewidth{0pt}
\tablecaption{Modeled fractional abundances of carbon-chain-oxide neutrals and anions \label{tab:cno}}
\tablehead{{Species}&{TMC-1}&{L1544}}
\startdata
C$_3$O&$2.31\times10^{-10}$&$5.77\times10^{-11}$\\
C$_6$O&$6.78\times10^{-11}$&$2.35\times10^{-12}$\\
C$_7$O&$1.33\times10^{-10}$&$1.02\times10^{-12}$\\
HC$_6$O&$5.81\times10^{-10}$&$2.58\times10^{-12}$\\
HC$_7$O&$1.46\times10^{-09}$&$2.16\times10^{-12}$\\
C$_6$O$^-$&$6.20\times10^{-12}$&$1.25\times10^{-13}$\\
C$_7$O$^-$&$2.08\times10^{-11}$&$1.16\times10^{-13}$\\
HC$_6$O$^-$&$7.38\times10^{-12}$&$2.79\times10^{-14}$\\
HC$_7$O$^-$&$1.45\times10^{-11}$&$1.25\times10^{-14}$
\enddata
\tablecomments{Fractional abundances (with respect to $n_{\rm H_2}$) are from Model {\sc iii}, with $n_{\rm H_2}=10^4$~cm$^{-3}$, $t=10^5$~yr for TMC-1, and $n_{\rm H_2}=10^6$~cm$^{-3}$, $t=10^4$~yr for L1544.}

\end{deluxetable}

\section{Temporal evolution of the anion-to-neutral ratio}
\label{sec:anr}

\begin{figure}
\begin{center}
\includegraphics[width=0.63\columnwidth,angle=270]{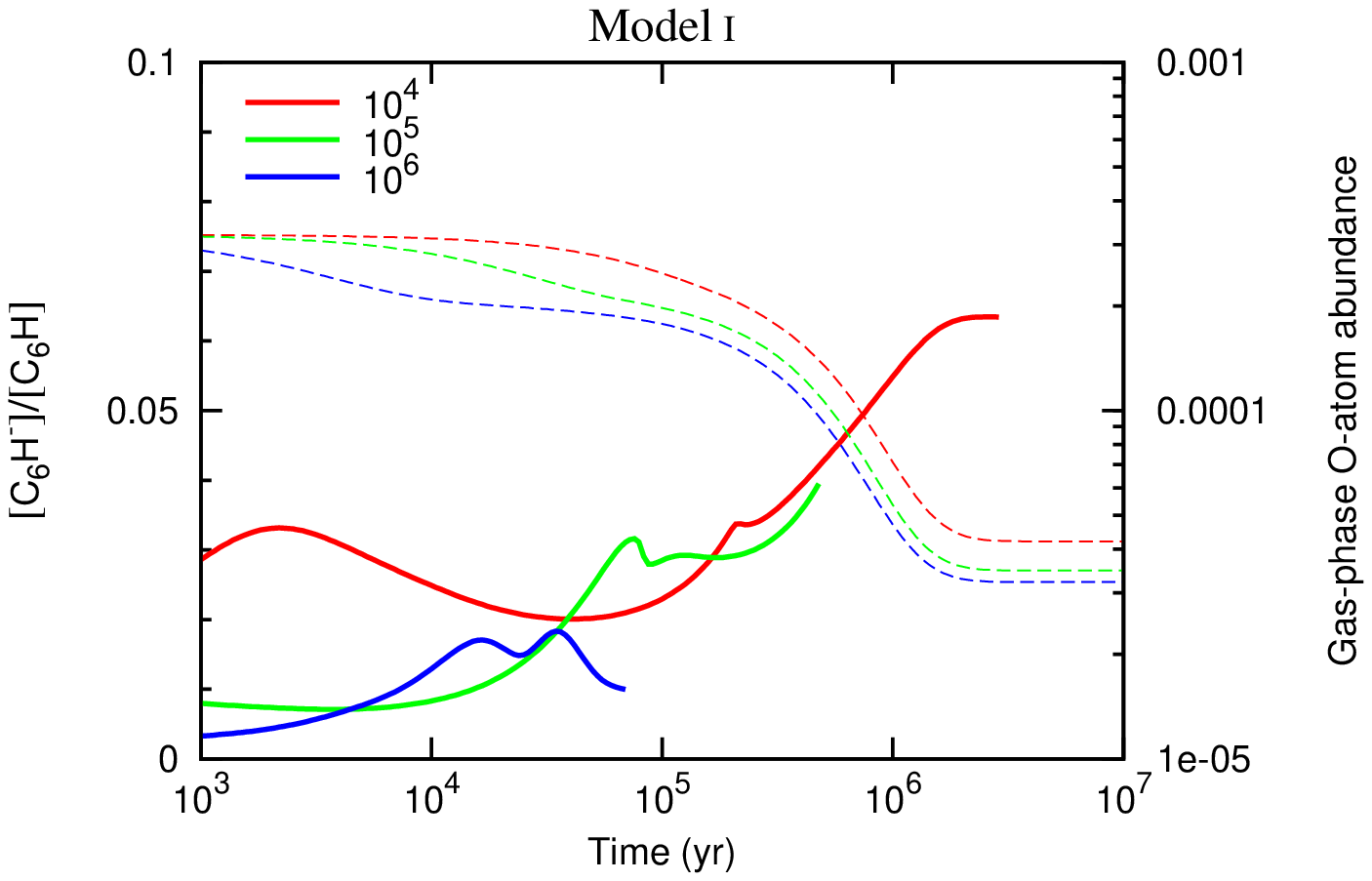}
\includegraphics[width=0.63\columnwidth,angle=270]{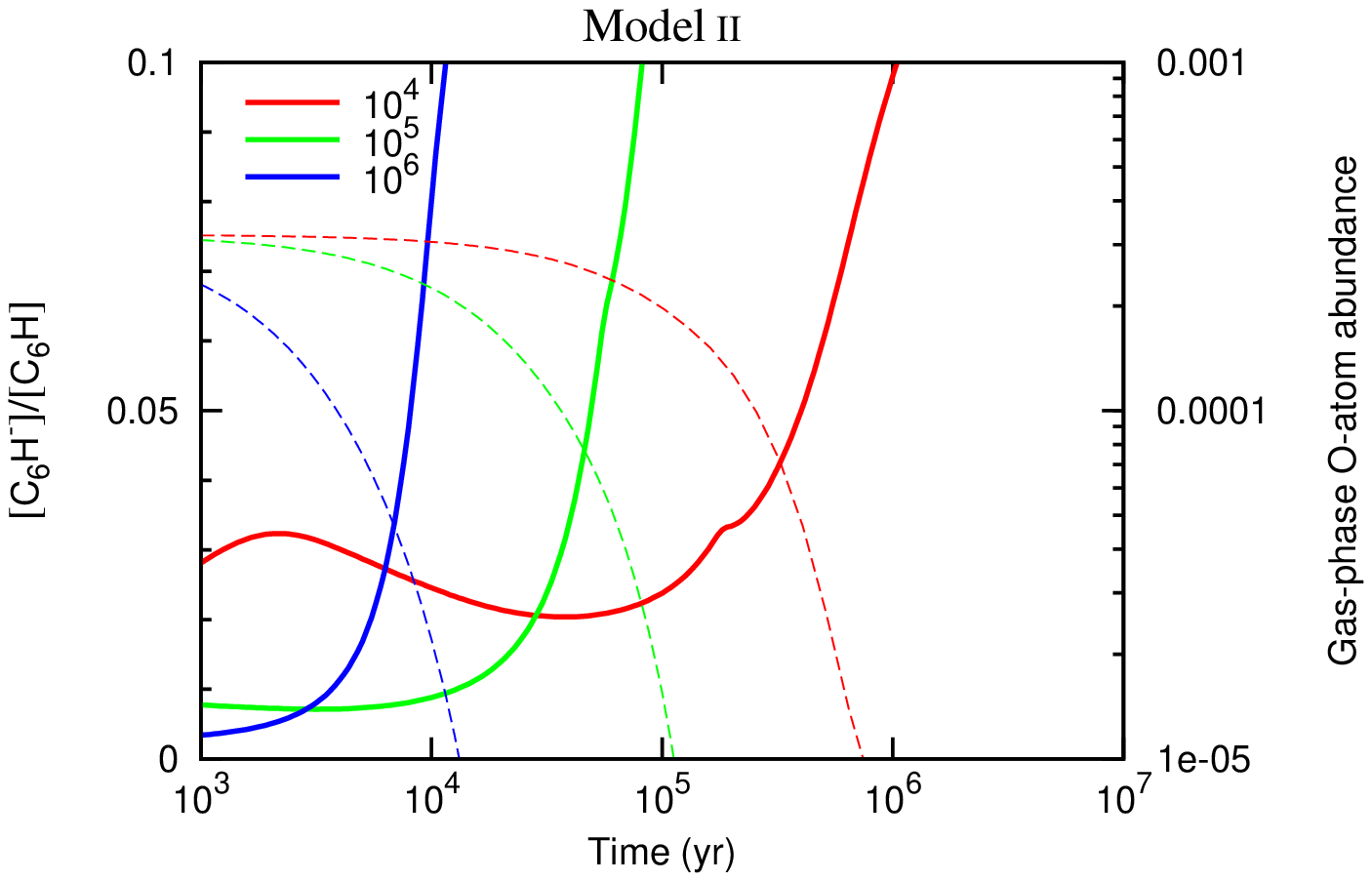}
\includegraphics[width=0.63\columnwidth,angle=270]{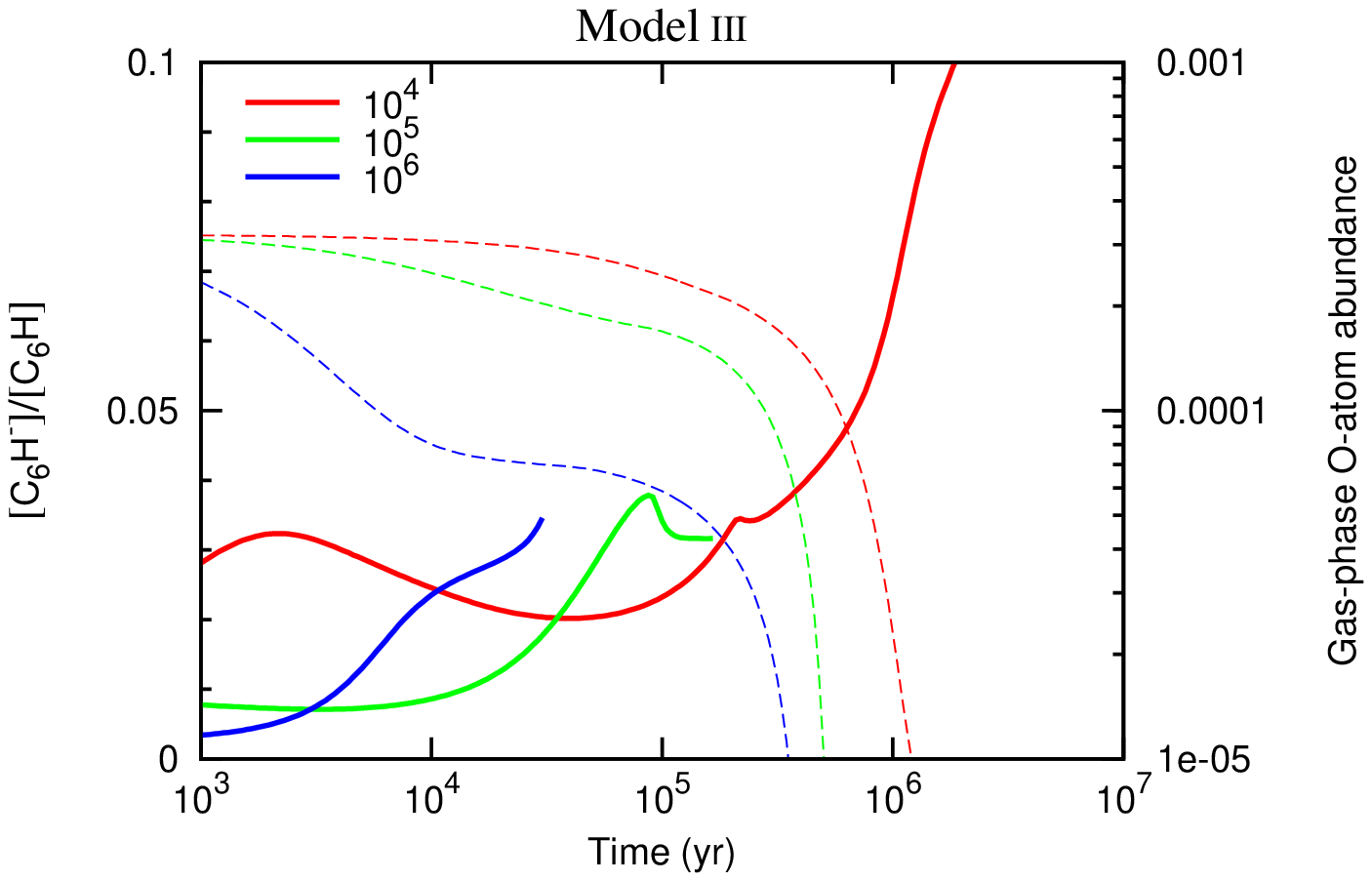}
\end{center}
\caption{Solid lines show modelled C$_6$H$^-$ anion-to-neutral ratios (left ordinate, linear scale) as a function of time for gas densities $n_{\rm H_2}= 10^4-10^6$ cm$^{-3}$ in the case of no accretion (Model {\sc i}; top), freeze-out accretion (Model {\sc ii}; middle) and accretion + CR-desorption (Model {\sc iii}; bottom). Curves have been truncated when neutral C$_6$H fractional abundances fall below $10^{-14}$. Dashed lines show the modelled gas-phase atomic oxygen fractional abundances (right ordinate, logarithmic scale).}
\label{fig:andens}
\end{figure}

Figure \ref{fig:andens} shows the behaviour of the C$_6$H$^-$ anion-to-neutral ratios as a function of time at H$_2$ densities of $10^4$, $10^5$ and $10^6$~cm$^{-3}$ (red, green and blue curves, respectively), for each of the three models ({\sc i}-{\sc iii}). The anion-to-neutral ratio patterns for C$_6$H$^-$ are representative of those for the other hydrocarbon anions in the model, and show clear variation as a function of cloud age and density. The anion-to-neutral ratio curves are distinctly different between the three models primarily as a result of their dependence on the gas-phase abundances of electrons and atomic H, C and O, which themselves vary with time and density. A common trend between the three models is that in denser media, faster cation recombination rates reduce the free electron abundances during the early stages of cloud evolution, resulting in lower anion production rates. At all modelled densities, the anion-to-neutral ratios tend to increase over time as H, C and O abundances fall and electron abundances rise.

For reference, at \n{6} our electron abundance as a function of time closely matches that shown in Figure 1 of \citet{rob03}. In our Model {\sc ii} (Figure \ref{fig:andens}, middle panel), the steep rise in the anion-to-neutral ratio is mainly a consequence of the rising electron abundance that occurs during the advanced stages of freeze-out. Dashed lines in Figure \ref{fig:andens} show the gas-phase \mbox{O-atom} fractional abundance, which may be used as an indicator for the level of depletion. We note that our models show qualitative agreement with the trends in the CO depletion levels ($\delta_{\rm CO}$) observed in TMC-1, L1512 and L1544; the CO goes from almost undepleted in TMC-1 to highly-depleted in the denser gas of L1544 (see Table \ref{tab:obs}).

Comparing Models {\sc i} and {\sc ii} (top and middle panels of Figure \ref{fig:andens}), the effects of gas-grain interactions are clear. Gas-phase species (especially lighter, atomic species), freeze out onto the dust grains at a faster rate in denser media, which has a strong impact on the anion abundances over time: C$_6$H$^-$ reacts quickly with atomic oxygen \citep{eic07}, so the depletion of O from the gas contributes towards a steady rise in [C$_6$H$^-$]/[C$_6$H]. As the depletion of gas-phase species proceeds towards total freeze-out -- indicated in the figures by the rapidly falling \mbox{O-atom} abundances -- electron densities rise by a factor of a few. This occurs primarily as a result of the falling abundances of H$_3$O$^+$ and HCO$^+$. These species have rapid recombination rates, so their loss results in elevated electron densities. Accordingly, the [C$_6$H$^-$]/[C$_6$H] ratio rises to $\sim50$\% towards the end of the cloud's lifetime (\emph{i.e.} towards the time of complete freeze-out). 

Such large anion-to-neutral ratios are incompatible with the interstellar anion observations made thus far (shown in Table \ref{tab:obs}). Given the steeply-rising nature of the [C$_6$H$^-$]/[C$_6$H] curves in Model {\sc ii} and the fact that the C$_6$H$^-$ abundance remains at an observable level ($\gtrsim10^{-13}$) until relatively late times, a broad range of anion-to-neutral ratios from approximately 0-50\% is predicted. However, all values observed so far have been less than 10\%. It seems statistically unlikely that observations have captured only the youngest clouds on the lower parts of these curves. 

Upon inclusion of cosmic-ray-induced desorption of atoms in Model {\sc iii} (Figure \ref{fig:andens}, bottom panel), the anion-to-neutral ratios in clouds with \n{5} and \n{6} obtain values of only a few percent, which are consistent with observations. Including desorption in these models keeps the electron density at a steady value of $\sim10^{-8}n_{\rm H_2}$ throughout the time of peak carbon-chain synthesis until well after the C~$\longrightarrow$~CO conversion has occurred. As a result, radiative electron attachment rates and hence anion abundances remain approximately constant (within a factor of a few) for most of the cloud's lifetime. At \n{4}, the results match observations at `early time' ($\sim10^5$~yr), but towards later times, [C$_6$H$^-$]/[C$_6$H] continues to rise (due to rising electron densities), and levels off at just over 10\%. This result implies that anion-to-neutral ratios greater than presently observed may be detectable in quiescent clouds significantly older than TMC-1. The decline of carbon-chain abundances at late times will make such observations challenging, however.

\section{Discussion}

The C$_6$H$^-$ anion-to-neutral ratio observed in L1527 (9.3\%) is significantly larger than in the other sources with similar densities (see Table \ref{tab:obs}). Accordingly, Model {\sc iii} gives a poorer agreement with L1527 than L1544 and L1521F. This discrepancy might be related to the presence of the low-mass protostar IRAS 04368+2557 at the center of L1527, heating from which is not accounted for by our model. Fits to the sub-mm SED have shown that the dust surrounding IRAS 04368+2557 has a temperature of 22-35 K.  The large abundances of carbon-chain-bearing molecules (including C$_n$H, C$_n$H$^-$, HC$_n$N and CH$_3$CCH) observed by \citet{sak08} have been interpreted as arising in the warm protostellar envelope as a result of `warm carbon-chain chemistry'. It has been theorized that in this region, methane ice sublimates from the dust grain surfaces and participates in a gas-phase chemistry which results in the production of large abundances of carbon-chain-bearing species. 

\citet{har08} modelled the gas-phase chemistry of L1527 using a similar approach to our Model {\sc i} (with \n{6}), but with initial abundances appropriate for a protostellar envelope at 30~K. Volatile molecular ices (including CO and CH$_4$) were theorized to have been formed on the dust, then released into the gas phase during protostellar warm-up. The large initial gas-phase methane abundance used by \citet{har08} results in greatly-accelerated synthesis of carbon-chain-bearing species as a result of ion-molecule chemistry \citep[see \emph{e.g.}][]{bro91,has08}, which explains why their modelled polyyne and cyanopolyyne abundances \citep[][Figure 5]{har08} are able to match observations after only a few thousand years. Our Models {\sc ii} and {\sc iii}, on the other hand, give a satisfactory agreement with this source after $\sim10^4$~yr, and show that warm carbon-chain chemistry is not necessarily required to explain the large carbon chain abundances in this source. Instead, the depletion of oxygen atoms onto dust may be sufficient to permit an efficient carbon chain chemistry to develop. The presence of abundant hydrocarbon anions is also likely to boost the abundances of carbon-chain-bearing species \citep[see][]{wal09}. However, warm-up chemistry may explain the elevated L1527 anion-to-neutral ratio if it can produce greater electron densities or reduced H and \mbox{O-atom} abundances; to test these ideas would require a more detailed chemical model. Indeed, \citet{har08} calculated a C$_6$H$^-$ anion-to-neutral ratio of about 50\% in L1527 (five times greater than observed), which is likely to be at least partly a result of their low initial H, C and \mbox{O-atom} abundances, which are are several orders of magnitude less than ours.

Agreement of our model with observations at $t\sim10^4$~yr is in apparent contradiction with the age of L1527, which is likely to be $\gtrsim10^5$~yr based on the estimated length of the pre-stellar core phase ($\sim10^5$~yr; \citealt{kir05}). However, the gas in the protostellar envelope has undergone collapse from lower density to arrive at the present value of \n{6}. Given the density-dependence of the timescale for formation of peak carbon chain abundances (see Figures \ref{fig:n4}-\ref{fig:n6}), our derived time-of-best-agreement cannot be treated as a reliable age indicator.

The adopted initial abundances for our models (Table \ref{tab:parents}) assume the cloud evolves from a neutral state at $t=0$. This is contrary to \citet{wal09} who assumed all the C and S atoms to be initially ionized. As can be seen by comparing the results of our \n{4} model with theirs, the choice of initial ionisation conditions has a profound impact on the neutral and anionic polyyne abundances at very early times ($t\lesssim500$~yr). The large abundance of free electrons in the `initially-ionized' model results in rapid anion production. Consequently, given the propensity of hydrocarbon anions to stimulate synthesis of enhanced polyyne and polyacetylene abundances \emph{via} anion-neutral reactions, the carbon-chain-bearing species can reach abundances orders of magnitude greater than in our `initially-neutral' models. The density-dependence of the chemical timescales (see Section \ref{sec:results}) determines whether this `very early-time' peak in carbon chain abundances has any lasting impact on the abundances of the species of interest.

For instance, at \n{4}, by the time of best fit with observations ($t\sim10^5$~yr), the abundances in the initially-neutral and initially-ionized models have converged so as to be practically indistinguishable. The same can also be said for the models at \n{5}. However, towards higher densities ($n\gtrsim10^6$~cm$^{-3}$), convergence does not occur sufficiently quickly, and the elevated carbon chain abundances persist through until the time of freeze-out. These effects are more pronounced for longer carbon chains, and as a result, the C$_6$H abundance is about a factor of two greater for \n{6} at around the time of best fit with observations ($t\sim10^4$~yr), when carbon is initially ionized. Anion-to-neutral ratios are also greater by a factor of about 0.5. Although still in good agreement with observations, it is deemed that models with C (and S) initially ionized are less representative of the conditions present in such dense clouds, which are generally found in regions of high extinction where ionizing UV photons cannot penetrate and the gas is therefore neutral.

Regarding the possible use of anions as chemical probes, the reaction rates of C$_6$H$^-$ with O and H atoms were measured in the laboratory by \citet{eic07} as $5.4\times10^{-10}$ and $5.0\times10^{-10}$~cm$^3$\,s$^{-1}$, respectively. Because of this similarity, the dominant C$_6$H$^-$ destruction mechanisms can be considered to be by reaction with H atoms for gas-phase [O]/[H]~$<1$ and by reaction with \mbox{O-atoms} for [O]/[H]~$>1$. Although we use a slightly larger rate for reactions of C$_6$H$^-$ with \mbox{C-atoms} ($1.0\times10^{-9}$~cm$^3$\,s$^{-1}$), the \mbox{C-atom} abundances remain sufficiently low in our models that this reaction is never a competitive cause of C$_6$H$^-$ destruction. The \mbox{N-atom} destruction reactions are also a couple of orders of magnitude less rapid.  At a density \n{4}, H is more abundant than O by a factor of at least 2.5 for all times in each of our models.  However, at higher densities ($n_{\rm H_2}\gtrsim10^5$~cm$^{-3}$) -- as a consequence of the more rapid gas-phase chemistry and H$_2$ formation on dust -- H becomes up to 2 orders of magnitude less abundant than O at early times. In these circumstances, at around the times of best agreement of the models with observations, the anion abundances are closely dependent on the \mbox{O-atom} abundances. Anion abundances are also sensitive to the number density of free electrons. Therefore, if the electron abundance is known, anion-to-neutral ratios may be used as a probe of the gas-phase oxygen abundance in dense interstellar clouds.

The absolute and relative carbon chain and anion abundances in our models are dependent on the gas accretion rates, which scale with the total dust grain surface area and the sticking coefficients, both of which are subject to uncertainty. To address this, we have run models employing a factor of four range in dust surface area (by taking $X_G'=2X_G$ and $X_G'=X_G/2$), and find that although the modelled abundances and times of best agreement with observations change slightly (greater differences occurring in the higher-density models), the overall results of our study are unaffected. Increased accretion rates ($X_G'=2X_G$) lead to a stronger, earlier freeze-out peak in the polyyne and anion abundances, combined with a steeper rise in the electron abundances and anion-to-neutral ratios at late times. Reduced accretion rates ($X_G'=X_G/2$), on the other hand, result in significantly greater late-time abundances, allowing complex molecules to exist in the gas for almost twice as long. Examination of the effects of species-dependent differential accretion rates, further to what may be inferred from our treatment of atomic desorption in Model {\sc iii}, is beyond the scope of the present study.

\section{Summary}

Similar to previous models for anions  in TMC-1 \citep{mil07,wal09}, for a density of \n{4} we obtain good agreement between modelled and observed anion-to-neutral ratios and absolute polyyne abundances at times $\sim10^5$~yr. Similarly good agreement is found for the quiescent cloud Lupus 1A. This result is independent of the inclusion of gas-grain interactions, which are negligible at this time and for this relatively low gas density. Models give a poor fit with observations at later times due to their falling cyanopolyyne abundances, although the freeze-out of atomic oxygen does permit the polyynes to remain abundant up to $\sim10^6$~yr.

The chemical timescale for the formation of abundant carbon-chain-bearing species (including the polyynes, cyanopolyynes and hydrocarbon anions) is inversely dependent on the cloud density. Thus, for clouds of moderate density (\n{5}), such as L1512 and L1251A, the best fits between modelled and observed abundances of carbon-chain-bearing species occur at earlier times ($\sim7\times10^4$~yr). At this higher density, Models {\sc i} (no accretion) and {\sc iii} (accretion + CR-desorption) match the observations better than Model {\sc ii} (accretion with no desorption).

At densities \n{6}, the best fit with observations of L1544, L1521F and L1527 is for Model {\sc iii}, at $t\sim10^4$~yr. We find that by accounting for gas-grain interactions (in particular, the freeze-out of atomic oxygen), the large observed abundances of carbon-chain-bearing species in these environments can be reproduced. However, a sufficient number of small O-bearing ions are required to remain in the gas-phase to maintain a steady electron abundance and therefore keep the radiative electron attachment rates down, in order to reproduce the relatively low observed anion-to-neutral ratios (on the order of a few percent) in L1544 and L1521F.  Without atomic desorption, rising electron abundances around the time of freeze-out result in excessively large anion-to-neutral ratios.

In summary, the gas-grain interaction, through its influence on the chemistry (in particular, the O and \mbox{H-atom} abundances and the electron density), is found to have profound effects on the abundances of carbon-chain-bearing species and their anions in dense interstellar clouds with $n_{\rm H_2}\gtrsim10^5$. Due to their destructive reactions with atomic oxygen, neutral and anionic polyynes are predicted to be most abundant in oxygen-depleted environments. In addition, \mbox{O-atom} + anion reactions are found to have the potential to produce as-yet unobserved interstellar species including carbon-chain oxides and C$_2$H$^-$. 

The sensitivity of the interstellar C$_6$H$^-$ anion-to-neutral ratio to the gas-phase O, H and $e^-$ abundances could allow anions to be used as a probe of these fundamental chemical species.

\acknowledgments
We acknowledge Dr. Catherine Walsh for assistance in development of the anion chemical network and Prof. Veronica Bierbaum and Dr. Zhibo Yang for sharing their laboratory data pertaining to reactions between oxygen atoms and anions. This research was supported NASA's Exobiology and Origins of Solar Systems programs and by the NASA Astrobiology Institute through the Goddard Center for Astrobiology.

\end{document}